\documentclass[aip, apl, reprint]{revtex4-1}

\pdfoutput=1 
\usepackage{graphicx}
\usepackage{color,amsmath}
\usepackage[normalem]{ulem}

\usepackage[colorlinks,linkcolor=blue,citecolor=blue,urlcolor=blue]{hyperref}
\usepackage{siunitx}
\usepackage{comment}
\usepackage{layouts}

\begin{document}
\title{Passivation of Edge States in Etched InAs Sidewalls}

\author{Christopher Mittag}
\email{mittag@phys.ethz.ch}
\affiliation{Solid State Physics Laboratory, Department of Physics, ETH Zurich, 8093 Zurich, Switzerland}

\author{Matija Karalic}
\affiliation{Solid State Physics Laboratory, Department of Physics, ETH Zurich, 8093 Zurich, Switzerland}

\author{Susanne M{\"u}ller}
\affiliation{Solid State Physics Laboratory, Department of Physics, ETH Zurich, 8093 Zurich, Switzerland}

\author{Thomas Tschirky}
\affiliation{Solid State Physics Laboratory, Department of Physics, ETH Zurich, 8093 Zurich, Switzerland}

\author{Werner Wegscheider}
\affiliation{Solid State Physics Laboratory, Department of Physics, ETH Zurich, 8093 Zurich, Switzerland}

\author{Olga Nazarenko}
\affiliation{Laboratory of Inorganic Chemistry, Department of Chemistry and Applied Biosciences, ETH Zurich, 8093 Zurich, Switzerland}

\author{Maksym V. Kovalenko}
\affiliation{Laboratory of Inorganic Chemistry, Department of Chemistry and Applied Biosciences, ETH Zurich, 8093 Zurich, Switzerland}

\author{Thomas Ihn}
\affiliation{Solid State Physics Laboratory, Department of Physics, ETH Zurich, 8093 Zurich, Switzerland}

\author{Klaus Ensslin}
\affiliation{Solid State Physics Laboratory, Department of Physics, ETH Zurich, 8093 Zurich, Switzerland}

\date{\today}

\begin{abstract}
We investigate different methods of passivating sidewalls of wet etched InAs heterostructures in order to suppress inherent edge conduction that is presumed to occur due to band bending at the surface leading to charge carrier accumulation. Passivation techniques including sulfur, positively charged compensation dopants and plasma enhanced chemical vapor deposition of $\mathrm{SiN}_{\mathrm{x}}$ do not show an improvement. Surprisingly, atomic layer deposition of $\mathrm{Al}_{\mathrm{2}}\mathrm{O}_{\mathrm{3}}$ leads to an increase in edge resistivity of more than an order of magnitude. While the mechanism behind this change is not fully understood, possible reasons are suggested.  
\end{abstract}

\maketitle

In contrast to the most commonly used materials in semiconductor physics, GaAs and Si, which feature a Fermi level pinned at mid-gap in the case of GaAs or a well-insulating native oxide layer in the case of Si, InAs shows a peculiar specialty. At the surface, its conduction band (CB) is bending downward and its Fermi level is pinned above the CB minimum, which leads to an electron accumulation.~\cite{noguchi_intrinsic_1991, olsson_charge_1996} Recently, a double quantum well structure containing InAs and GaSb has been suggested as a quantum spin Hall insulator (QSHI)~\cite{liu_quantum_2008} and edge transport in etched Hall bars has been verified.~\cite{knez_evidence_2011,suzuki_edge_2013,mueller_nonlocal_2015} However, edge transport of a similar magnitude was also found in InAs/GaSb samples which are not in the topological phase.~\cite{nichele_edge_2016,nguyen_decoupling_2016} The presence of this trivial edge transport and the aforementioned accumulation of charge carriers at InAs surfaces suggests that it originates from the InAs layer and is verified by observing edge transport in single InAs QWs.~\cite{mueller_edge_2017} This unwanted edge conduction can not only obscure data in the intricate InAs/GaSb material system, but is also detrimental to realizing quantum devices in InAs heterostructures, as mesa edges and structures defined by etching introduce a parallel conducting channel, and InAs nanowires, where surface accumulation is believed to drastically lower the mobility at low temperatures.~\cite{pfund_fabrication_2006, tilburg_surface_2010} In recent work, these problems could be controlled for field effect transistors (FETs), displaying promising room temperature operation of devices in industry dimensions.~\cite{punkkinen_oxidized_2011,wang_inas_2013,oxland_inas_2016}

\begin{figure}[!b]
	\includegraphics[width=\columnwidth]{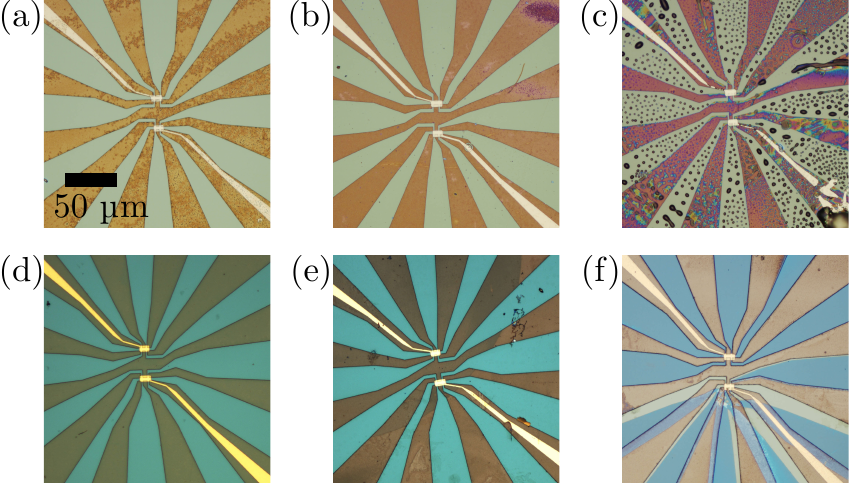}
	\caption{Optical microscope images of a selection of samples, displaying visual changes of different passivation techniques employed: \textbf{(a)} no treatment after etching, deposition of $150\,\mathrm{nm}$ $\mathrm{SiN}_{\mathrm{x}}$ by PECVD \textbf{(b)} immersion in $\mathrm{Mg}(\mathrm{BH}_4)_2$, subsequent $\mathrm{SiN}_{\mathrm{x}}$ deposition  \textbf{(c)} treatment with PCBM, subsequent $\mathrm{SiN}_{\mathrm{x}}$ deposition \textbf{(d)} no treatment after etching, deposition of $40\,\mathrm{nm}$ $\mathrm{Al}_{\mathrm{2}}\mathrm{O}_{\mathrm{3}}$ by ALD \textbf{(e)} immersion in TAM followed by $\mathrm{Al}_{\mathrm{2}}\mathrm{O}_{\mathrm{3}}$ deposition \textbf{(f)} immersion in ammonium sulfide followed by $\mathrm{Al}_{\mathrm{2}}\mathrm{O}_{\mathrm{3}}$ deposition. The scale bar applies to all images \textbf{(a)}-\textbf{(f)}.} 
	\label{fig1}
\end{figure} 

In this paper, we investigate InAs Hall bars which display edge conduction after the bulk is pinched off by top-gating and apply different passivation techniques during fabrication with the goal of alleviating this edge conduction. Our heterostructure is grown by molecular beam epitaxy (MBE) on a GaAs substrate and consists of $200\,\mathrm{nm}$ GaAs, followed by $6\,\mathrm{nm}$ AlAs, a $32\,\mathrm{nm}$ AlSb layer, a $1100\,\mathrm{nm}$ Al$_{\mathrm{x}}$Ga$_{1-\mathrm{x}}$Sb layer ($\mathrm{x}=\,65\%$), a $500\,\mathrm{nm}$ GaSb layer, a $50\,\mathrm{nm}$ superlattice of 10 iterations of $2.5\,\mathrm{nm}$ AlSb and $2.5\,\mathrm{nm}$ GaSb layers and then the $15\,\mathrm{nm}$ InAs QW sandwiched by two $50\,\mathrm{nm}$ AlSb barriers and a $3\,\mathrm{nm}$ GaSb capping layer to prevent oxidation of the barrier. Ohmic contacts (Ge/Au/Ni/Au) have been evaporated and $4\times 8\,\mathrm{\mu m}$ Hall bars were defined using standard wet etch recipes.~\cite{chaghi_wet_2009,pal_influence_2015} Following the etching process and resist removal in acetone and isopropanol, different passivation techniques have been applied, which consisted either of a chemical dip and dielectric deposition, or the direct deposition of a dielectric. This separates the structure from the subsequently evaporated Ti/Au top gate. As InAs/GaSb heterostructures have widespread use in infrared detection where surface accumulation can lead to unwanted signals, different passivation techniques have been explored for optoelectronic devices.~\cite{plis_passivation_2013} A range of these are applied to transport measurements in the following. The most widespread technique is sulfur passivation, using either ammonium sulfide~\cite{petrovykh_chemical_2003}, thioacetamide (TAM)~\cite{petrovykh_surface_2005} or 1-octadecanethiol (ODT).~\cite{hang_role_2008,sun_removal_2012} The general idea behind this method is the creation of covalent bonds to a layer of sulfur adatoms, reducing the density of surface states. For the ammonium sulfide passivations, samples are dipped in a $20\,\%\,(\mathrm{NH}_4)_2\mathrm{S}$ solution at room temperature for up to 15 minutes. As can be seen in Fig.~\ref{fig1}(f), this chemical is too aggressive and etches parts of the heterostructure, presumably the AlSb barriers. This leads to visible "lifting off" of parts of the mesa structure and results in nonfunctional devices whose transport behavior does not depend on gate voltages and magnetic fields in a way characteristic for two-dimensional electron gases. Reducing the dip time and diluting the solution further (up to $1:4$) in $\mathrm{H}_2\mathrm{O}$ did not resolve this problem, which renders ammonium sulfide unsuitable for our purpose. A second, more gentle form of sulfur passivation is investigated by using TAM in $\mathrm{H}_2\mathrm{O}$ at a concentration of $0.2~\mathrm{mol}/\mathrm{l}$, at a temperature of $60^{\circ}\mathrm{C}$ and at a pH of $2$ which was adjusted by adding acetic acid, following Petrovykh et al.~\cite{petrovykh_surface_2005} The time of the dip was again varied up to 15 minutes and did neither influence the resulting data nor the visual appearance of the sample, compare Fig.~\ref{fig1}(e). Additionally, we explore two other methods of passivation in magnesium borohydride ($\mathrm{Mg}(\mathrm{BH}_4)_2$) and phenyl-C61-butyric acid methyl ester (PCBM). The $\mathrm{Mg}(\mathrm{BH}_4)_2$ is intended to reduce native and other oxides after wet etching and to then terminate the surface with $\mathrm{Mg}^{2+}$, thereby acting as a "compensation doping" for the surplus of negative charge carriers at the surface. PCBM, which is a known electron acceptor in organic photovoltaics, could also act as an acceptor for these charges. $\mathrm{Mg}(\mathrm{BH}_4)_2$ was dissolved in tetrahydrofuran (THF) in a glovebox at argon atmosphere, into which the samples had been introduced before. The samples were then dipped into the solution for two minutes at room temperature and at $60^{\circ}\mathrm{C}$ respectively. No influence of the temperature could be determined. This treatment also preserved the sample quality, as can be seen in Fig.~\ref{fig1}(b). Performing a dip was not possible for PCBM. Therefore, using a pipette, a toluene solution of PCBM was dropped onto the sample lying on a hotplate in the glovebox. This technique unfortunately resulted in very uneven spatial distribution and the formation of small droplets, which can be seen in Fig.~\ref{fig1}(c). They made subsequent cleanroom fabrication unfeasible. The sample shown featured the smallest amount dropped ($5\,\mathrm{\mu L}$), and was the only one where a working gate could be fabricated, albeit with very limited tuning range, as will be discussed later. Two other kinds of samples have not been chemically passivated, but were solely overgrown with $150\,\mathrm{nm}$ of $\mathrm{SiN}_{\mathrm{x}}$ by plasma enhanced chemical vapor deposition (PECVD) at a temperature of $300\,^{\circ}\mathrm{C}$ or $40\,\mathrm{nm}$ of $\mathrm{Al}_{\mathrm{2}}\mathrm{O}_{\mathrm{3}}$ by atomic layer deposition (ALD) at a temperature of $150\,^{\circ}\mathrm{C}$. Another sample was etched and then left at cleanroom air atmosphere for two weeks before continuing fabrication (with an ALD $\mathrm{Al}_{\mathrm{2}}\mathrm{O}_{\mathrm{3}}$ gate dielectric) in order to investigate oxidized edges. They are believed to be conducting, although experimental data on their transport behavior has not been reported in the literature so far.

\begin{figure}[!t]
	\includegraphics[width=\columnwidth]{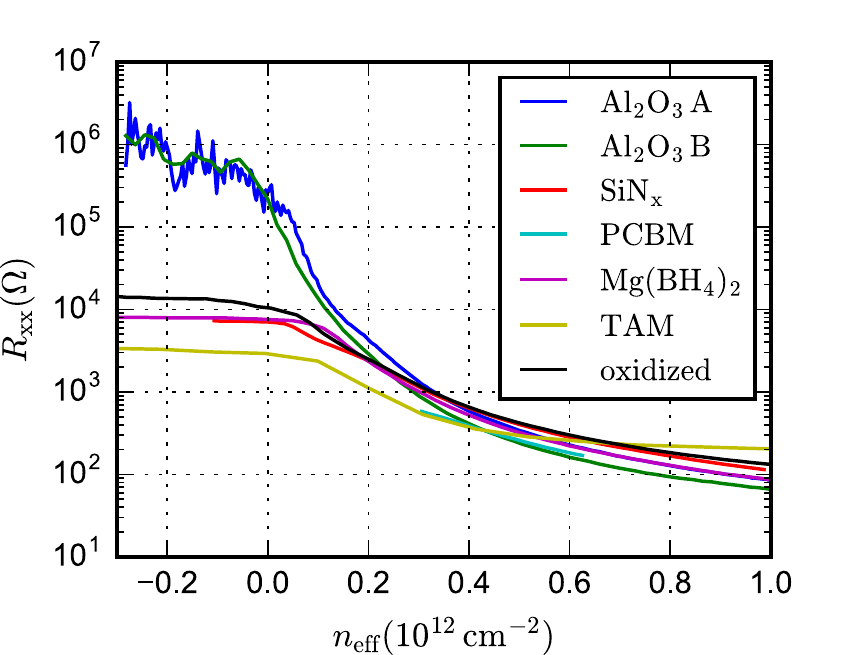}
	\caption{Longitudinal resistances $R_\mathrm{xx}$ of Hall bars passivated by different techniques as a function of effective density $n_{\mathrm{eff}}$, obtained by extending a linear fit of the gate voltage to density characteristic to negative voltages. Positive values of this quantity correspond to the density in the quantum well, while negative values provide a renormalized gate voltage axis allowing comparison of devices with different gate dielectrics and thicknesses. ALD $\mathrm{Al}_{\mathrm{2}}\mathrm{O}_{\mathrm{3}}$ samples shown are passivated directly after wet etching (A) and with one hour of waiting time (B).}  
	\label{fig2}
\end{figure}

In Fig.~\ref{fig2} we show transport data comparing the previously described methods of passivation. The longitudinal resistance $R_\mathrm{xx}$ of Hall bars, measured by low frequency AC lock-in techniques at $T=1.3\,\mathrm{K}$, is plotted as a function of effective density $n_{\mathrm{eff}}$. This quantity has been calculated by determining the relation between electron density $n$ in the QW and the top gate voltage, $V_{\mathrm{tg}}$, from Hall effect measurements, and extending the scale linearly to negative gate voltages. We want to emphasize that only $n_{\mathrm{eff}}\geq 0$ has a physical meaning corresponding to the density in the QW and negative values merely provide a renormalized gate voltages axis which is independent of different capacitances and charge redistributions. From Corbino disk measurements in InAs it is known that the QW can be completely depleted by a top gate, driving the resistance in the bulk of the device to arbitrarily high values, whereas for Hall bars with etched edges it saturates in the $\mathrm{k\Omega}$ range.~\cite{mueller_edge_2017} A successful passivation is therefore expected to either completely suppress the edge conduction, facilitating insulating behavior after QW pinch-off, or at least increase their resistance to a value high enough to unambiguously observe ballistic transport phenomena in InAs-based material systems, i.e. $R_{\mathrm{edge}}\gg\frac{\mathrm{h}}{\mathrm{e}^2}$ for typical edge lengths. Turning to Fig.~\ref{fig2}, we can immediately distinguish two different families of curves. For samples passivated with TAM, $\mathrm{Mg}(\mathrm{BH}_4)_2$, $\mathrm{SiN}_{\mathrm{x}}$ and the oxidized sample, $R_\mathrm{xx}$ saturates between $3-14\,\mathrm{k\Omega}$. This allows us to draw the following conclusions: firstly, our measurements are consistent with the results of Ref.~\citenum{mueller_edge_2017}, where edge conductance of the same order of magnitude was measured in PECVD $\mathrm{SiN}_{\mathrm{x}}$ overgrown Hall bars. Secondly, the presumption that oxidized samples display edge conduction is correct, and its magnitude is similar to samples which have been overgrown by PECVD directly after etching, and therefore have not been purposely oxidized. Here it is important to mention that the exact dynamics of the oxidation process are not well understood, which means that edges could already start to oxidize during resist removal necessary after wet etching and the transfer time to the vacuum chamber for dielectric deposition. Thirdly, neither the sulfur passivation with TAM nor the $\mathrm{Mg}(\mathrm{BH}_4)_2$ dip led to a significant change in transport behavior after pinch-off.

The two types of samples fabricated without chemical passivation and ALD $\mathrm{Al}_{\mathrm{2}}\mathrm{O}_{\mathrm{3}}$ display a resistance increasing to $\mathrm{M\Omega}$ after depletion of the QW, which is surprising as one would naively expect the water precursor of the ALD process to further edge oxidation. After pinch-off, the resistance slightly increases with decreasing gate voltage and develops more noise. These fluctuations do not reproduce exactly in subsequent measurements. They lead us to believe that the carrier accumulation at the edge is sufficiently reduced such that transport is close to breaking down. Sample B, which was left in air for approximately one hour during chemical passivation of other samples and was intended as a reference sample, surprisingly first showed enhanced resistance. After this observation, the process was repeated without waiting time (sample A), displaying a further increase in resistance. Therefore, we recognize that the oxidation process is happening on timescales slower than one hour, but faster than two weeks (c.f. oxidized sample, black curve in Fig.~\ref{fig2}). Finally, the data of the PCBM passivated sample (cyan curve in Fig.~\ref{fig2}) shows a tuning range limited by gate leakage which occured due to the severe fabrication difficulties described previously. Out of four devices, only one functioned, rendering PCBM unsuitable as a method of passivation. While the data in Fig.~\ref{fig2} show the resistance across the whole device, it has been verified by nonlocal transport measurements that the current flows along the edge for gate voltages below pinch-off, and that there are no other bulk leakage paths that the current can take after the QW is pinched off.

\begin{figure}[!t]
	\includegraphics[width=\columnwidth]{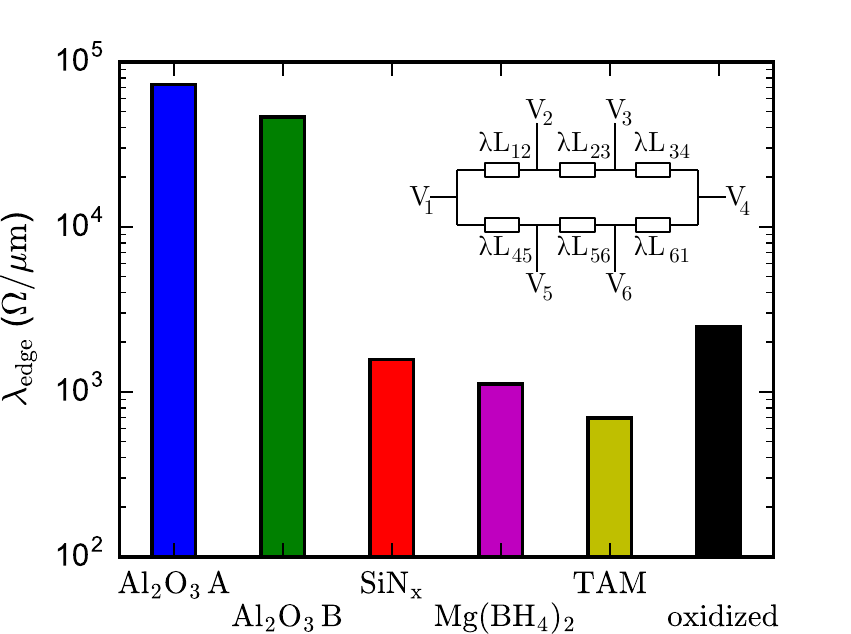}
	\caption{Bar chart comparing specific edge resistivities $\lambda_{\mathrm{edge}}$ of different passivation techniques. All methods investigated in this paper display $\lambda_{\mathrm{edge}} < 2.5\,\mathrm{k}\Omega$, with the exception of ALD $\mathrm{Al}_{\mathrm{2}}\mathrm{O}_{\mathrm{3}}$, where $\lambda_{\mathrm{edge}}$ reaches $72.8\,\mathrm{k}\Omega$. Inset: the resistor network model used to calculate the specific edge resistivity.}  
	\label{fig3}
\end{figure}

Using a resistor network model shown in the inset of Fig.~\ref{fig3}, it is possible to calculate a specific edge resistivity $\lambda_{\mathrm{edge}}$. The assumptions in this model are a complete QW pinch-off, current flowing entirely along the edges and an edge resistance that scales linearly with edge length. As previously mentioned, the first two assumptions are justified by Corbino disk and nonlocal measurements, whereas the last one is expected for a diffusive conduction path along the edge and is confirmed in our measurements. The specific value of $\lambda_{\mathrm{edge}}$ depends on the point chosen for analysis in the region where the QW is pinched off. This is because the edges can be gated weakly, resulting in a finite slope after depletion of the QW, which can be seen in Fig.~\ref{fig2}. For consistency we have evaluated the edge resistivity at the point $n_{\mathrm{eff}}=0$ for all samples.  The comparison of $\lambda_{\mathrm{edge}}$ for all feasible passivation techniques used is displayed in Fig.~\ref{fig3}, where $\lambda_{\mathrm{edge}}$ of the ALD $\mathrm{Al}_{\mathrm{2}}\mathrm{O}_{\mathrm{3}}$ passivation exceeds all other passivation techniques by more than an order of magnitude. 

The role of the ALD $\mathrm{Al}_{\mathrm{2}}\mathrm{O}_{\mathrm{3}}$ process in controlling the surface states of etched InAs heterostructures can be rationalized as follows. The trimethylaluminum, $\mathrm{Al}(\mathrm{CH}_{\mathrm{3}})_{\mathrm{3}}$, precursor used in the ALD process can act as a powerful reducing agent, which is known as  “self-cleaning”.~\cite{pi_semiconductor-insulator_1999} As an example, $\mathrm{Al}(\mathrm{CH}_{\mathrm{3}})_{\mathrm{3}}$ has been shown to remove native oxides from the surface of GaAs.~\cite{lee_ald_2010, tallarida_surface_2011,konda_effect_2011} There are also reports on surface reduction of InAs and InAs/GaSb heterostructures with $\mathrm{Al}(\mathrm{CH}_{\mathrm{3}})_{\mathrm{3}}$.~\cite{trinh_experimental_2011,hollinger_oxides_1994,timm_reduction_2010,salihoglu_atomic_2012} Reduction of native oxides is essentially already complete after the first $\mathrm{Al}(\mathrm{CH}_{\mathrm{3}})_{\mathrm{3}}$ pulse in the ALD process.~\cite{trinh_effects_2010} After the native oxides are reduced, only $\mathrm{Al}_{\mathrm{2}}\mathrm{O}_{\mathrm{3}}$ forms as it has a much lower Gibbs free energy of formation ($-377.9\,\mathrm{kcal}/\mathrm{mol}$)~\cite{pulver_thermal_2001} as compared to all other possible oxides on the surface of InAs (the Gibbs free energies of $\mathrm{In}_{\mathrm{2}}\mathrm{O}_{\mathrm{3}}$, $\mathrm{As}_{\mathrm{2}}\mathrm{O}_{\mathrm{3}}$, $\mathrm{As}_{\mathrm{2}}\mathrm{O}_{\mathrm{5}}$ are $-198.6\,\mathrm{kcal}/\mathrm{mol}$, $-137.7\,\mathrm{kcal}/\mathrm{mol}$, $-187.0\,\mathrm{kcal}/\mathrm{mol}$).~\cite{hollinger_oxides_1994} This strong chemical reducing capability of $\mathrm{Al}(\mathrm{CH}_{\mathrm{3}})_{\mathrm{3}}$ under ALD-conditions has been reported also for other systems.~\cite{juppo_trimethylaluminum_2001,alen_atomic_2001}

A possible reason for the ineffectiveness of sulfur passivation could be the fact that, as a group VI element, it provides an additional electron and can act as a donor, which in a different context has been reported to increase band bending at the surface.~\cite{lowe_extreme_2002,lowe_sulphur-induced_2003} This is also true for the related group VI element selenium.~\cite{watanabe_anomalous_1997} For transport applications, this effect seems to outweigh the advantages of terminating etched surfaces reported in optoelectronics.~\cite{plis_passivation_2013} This motivated the use of $\mathrm{Mg}(\mathrm{BH}_4)_2$ and PCBM to counter-dope at the surface, which, disregarding fabrication difficulties for PCBM, was not effective. Possible reasons could be too low or too high concentrations or usage of an unsuitable chemical, and can not be narrowed down at this point.

In conclusion, different methods to passivate edge states in transport through etched InAs QW structures have been investigated, of which ALD $\mathrm{Al}_{\mathrm{2}}\mathrm{O}_{\mathrm{3}}$ has been determined as the most effective technique, with a specific edge resistivity of up to $\lambda_{\mathrm{edge}}=72.8\,\mathrm{k}\Omega/\mathrm{\mu m}$. While edge conductance could not be removed completely, we hope that our findings are sufficient to stimulate research and applications in InAs and InAs-based materials and contribute to discussions on edge states in these systems. As an outlook for further efforts in passivation, finding a suitable chemical is still an open question. Another feasible option could be the passivation by MBE overgrowth with a related material of larger band gap, such as a quaternary alloy.~\cite{rehm_passivation_2005} 

%
   
\begin{acknowledgments}
The authors acknowledge the support of the ETH FIRST laboratory and the financial support of the Swiss Science Foundation (Schweizerischer Nationalfonds, NCCR QSIT).
\end{acknowledgments}

\bibliography{bibl}

\begin{thebibliography}{37}%
\makeatletter
\providecommand \@ifxundefined [1]{%
 \@ifx{#1\undefined}
}%
\providecommand \@ifnum [1]{%
 \ifnum #1\expandafter \@firstoftwo
 \else \expandafter \@secondoftwo
 \fi
}%
\providecommand \@ifx [1]{%
 \ifx #1\expandafter \@firstoftwo
 \else \expandafter \@secondoftwo
 \fi
}%
\providecommand \natexlab [1]{#1}%
\providecommand \enquote  [1]{``#1''}%
\providecommand \bibnamefont  [1]{#1}%
\providecommand \bibfnamefont [1]{#1}%
\providecommand \citenamefont [1]{#1}%
\providecommand \href@noop [0]{\@secondoftwo}%
\providecommand \href [0]{\begingroup \@sanitize@url \@href}%
\providecommand \@href[1]{\@@startlink{#1}\@@href}%
\providecommand \@@href[1]{\endgroup#1\@@endlink}%
\providecommand \@sanitize@url [0]{\catcode `\\12\catcode `\$12\catcode
  `\&12\catcode `\#12\catcode `\^12\catcode `\_12\catcode `\%12\relax}%
\providecommand \@@startlink[1]{}%
\providecommand \@@endlink[0]{}%
\providecommand \url  [0]{\begingroup\@sanitize@url \@url }%
\providecommand \@url [1]{\endgroup\@href {#1}{\urlprefix }}%
\providecommand \urlprefix  [0]{URL }%
\providecommand \Eprint [0]{\href }%
\providecommand \doibase [0]{http://dx.doi.org/}%
\providecommand \selectlanguage [0]{\@gobble}%
\providecommand \bibinfo  [0]{\@secondoftwo}%
\providecommand \bibfield  [0]{\@secondoftwo}%
\providecommand \translation [1]{[#1]}%
\providecommand \BibitemOpen [0]{}%
\providecommand \bibitemStop [0]{}%
\providecommand \bibitemNoStop [0]{.\EOS\space}%
\providecommand \EOS [0]{\spacefactor3000\relax}%
\providecommand \BibitemShut  [1]{\csname bibitem#1\endcsname}%
\let\auto@bib@innerbib\@empty
\bibitem [{\citenamefont {Noguchi}, \citenamefont {Hirakawa},\ and\
  \citenamefont {Ikoma}(1991)}]{noguchi_intrinsic_1991}%
  \BibitemOpen
  \bibfield  {author} {\bibinfo {author} {\bibfnamefont {M.}~\bibnamefont
  {Noguchi}}, \bibinfo {author} {\bibfnamefont {K.}~\bibnamefont {Hirakawa}}, \
  and\ \bibinfo {author} {\bibfnamefont {T.}~\bibnamefont {Ikoma}},\ }\href
  {\doibase 10.1103/PhysRevLett.66.2243} {\bibfield  {journal} {\bibinfo
  {journal} {Physical Review Letters}\ }\textbf {\bibinfo {volume} {66}},\
  \bibinfo {pages} {2243} (\bibinfo {year} {1991})}\BibitemShut {NoStop}%
\bibitem [{\citenamefont {Olsson}\ \emph {et~al.}(1996)\citenamefont {Olsson},
  \citenamefont {Andersson}, \citenamefont {H{\aa}kansson}, \citenamefont
  {Kanski}, \citenamefont {Ilver},\ and\ \citenamefont
  {Karlsson}}]{olsson_charge_1996}%
  \BibitemOpen
  \bibfield  {author} {\bibinfo {author} {\bibfnamefont {L.~{\"O}.}\
  \bibnamefont {Olsson}}, \bibinfo {author} {\bibfnamefont {C.~B.~M.}\
  \bibnamefont {Andersson}}, \bibinfo {author} {\bibfnamefont {M.~C.}\
  \bibnamefont {H{\aa}kansson}}, \bibinfo {author} {\bibfnamefont
  {J.}~\bibnamefont {Kanski}}, \bibinfo {author} {\bibfnamefont
  {L.}~\bibnamefont {Ilver}}, \ and\ \bibinfo {author} {\bibfnamefont {U.~O.}\
  \bibnamefont {Karlsson}},\ }\href {\doibase 10.1103/PhysRevLett.76.3626}
  {\bibfield  {journal} {\bibinfo  {journal} {Physical Review Letters}\
  }\textbf {\bibinfo {volume} {76}},\ \bibinfo {pages} {3626} (\bibinfo {year}
  {1996})}\BibitemShut {NoStop}%
\bibitem [{\citenamefont {Liu}\ \emph {et~al.}(2008)\citenamefont {Liu},
  \citenamefont {Hughes}, \citenamefont {Qi}, \citenamefont {Wang},\ and\
  \citenamefont {Zhang}}]{liu_quantum_2008}%
  \BibitemOpen
  \bibfield  {author} {\bibinfo {author} {\bibfnamefont {C.}~\bibnamefont
  {Liu}}, \bibinfo {author} {\bibfnamefont {T.~L.}\ \bibnamefont {Hughes}},
  \bibinfo {author} {\bibfnamefont {X.-L.}\ \bibnamefont {Qi}}, \bibinfo
  {author} {\bibfnamefont {K.}~\bibnamefont {Wang}}, \ and\ \bibinfo {author}
  {\bibfnamefont {S.-C.}\ \bibnamefont {Zhang}},\ }\href {\doibase
  10.1103/PhysRevLett.100.236601} {\bibfield  {journal} {\bibinfo  {journal}
  {Physical Review Letters}\ }\textbf {\bibinfo {volume} {100}},\ \bibinfo
  {pages} {236601} (\bibinfo {year} {2008})}\BibitemShut {NoStop}%
\bibitem [{\citenamefont {Knez}, \citenamefont {Du},\ and\ \citenamefont
  {Sullivan}(2011)}]{knez_evidence_2011}%
  \BibitemOpen
  \bibfield  {author} {\bibinfo {author} {\bibfnamefont {I.}~\bibnamefont
  {Knez}}, \bibinfo {author} {\bibfnamefont {R.-R.}\ \bibnamefont {Du}}, \ and\
  \bibinfo {author} {\bibfnamefont {G.}~\bibnamefont {Sullivan}},\ }\href
  {\doibase 10.1103/PhysRevLett.107.136603} {\bibfield  {journal} {\bibinfo
  {journal} {Physical Review Letters}\ }\textbf {\bibinfo {volume} {107}},\
  \bibinfo {pages} {136603} (\bibinfo {year} {2011})}\BibitemShut {NoStop}%
\bibitem [{\citenamefont {Suzuki}\ \emph {et~al.}(2013)\citenamefont {Suzuki},
  \citenamefont {Harada}, \citenamefont {Onomitsu},\ and\ \citenamefont
  {Muraki}}]{suzuki_edge_2013}%
  \BibitemOpen
  \bibfield  {author} {\bibinfo {author} {\bibfnamefont {K.}~\bibnamefont
  {Suzuki}}, \bibinfo {author} {\bibfnamefont {Y.}~\bibnamefont {Harada}},
  \bibinfo {author} {\bibfnamefont {K.}~\bibnamefont {Onomitsu}}, \ and\
  \bibinfo {author} {\bibfnamefont {K.}~\bibnamefont {Muraki}},\ }\href
  {\doibase 10.1103/PhysRevB.87.235311} {\bibfield  {journal} {\bibinfo
  {journal} {Physical Review B}\ }\textbf {\bibinfo {volume} {87}},\ \bibinfo
  {pages} {235311} (\bibinfo {year} {2013})}\BibitemShut {NoStop}%
\bibitem [{\citenamefont {Mueller}\ \emph {et~al.}(2015)\citenamefont
  {Mueller}, \citenamefont {Pal}, \citenamefont {Karalic}, \citenamefont
  {Tschirky}, \citenamefont {Charpentier}, \citenamefont {Wegscheider},
  \citenamefont {Ensslin},\ and\ \citenamefont {Ihn}}]{mueller_nonlocal_2015}%
  \BibitemOpen
  \bibfield  {author} {\bibinfo {author} {\bibfnamefont {S.}~\bibnamefont
  {Mueller}}, \bibinfo {author} {\bibfnamefont {A.~N.}\ \bibnamefont {Pal}},
  \bibinfo {author} {\bibfnamefont {M.}~\bibnamefont {Karalic}}, \bibinfo
  {author} {\bibfnamefont {T.}~\bibnamefont {Tschirky}}, \bibinfo {author}
  {\bibfnamefont {C.}~\bibnamefont {Charpentier}}, \bibinfo {author}
  {\bibfnamefont {W.}~\bibnamefont {Wegscheider}}, \bibinfo {author}
  {\bibfnamefont {K.}~\bibnamefont {Ensslin}}, \ and\ \bibinfo {author}
  {\bibfnamefont {T.}~\bibnamefont {Ihn}},\ }\href {\doibase
  10.1103/PhysRevB.92.081303} {\bibfield  {journal} {\bibinfo  {journal}
  {Physical Review B}\ }\textbf {\bibinfo {volume} {92}},\ \bibinfo {pages}
  {081303} (\bibinfo {year} {2015})}\BibitemShut {NoStop}%
\bibitem [{\citenamefont {Nichele}\ \emph {et~al.}(2016)\citenamefont
  {Nichele}, \citenamefont {Suominen}, \citenamefont {Kjaergaard},
  \citenamefont {Marcus}, \citenamefont {Sajadi}, \citenamefont {Folk},
  \citenamefont {Qu}, \citenamefont {Beukman}, \citenamefont {Vries},
  \citenamefont {Veen}, \citenamefont {Nadj-Perge}, \citenamefont
  {Kouwenhoven}, \citenamefont {Nguyen}, \citenamefont {Kiselev}, \citenamefont
  {Yi}, \citenamefont {Sokolich}, \citenamefont {Manfra}, \citenamefont
  {Spanton},\ and\ \citenamefont {Moler}}]{nichele_edge_2016}%
  \BibitemOpen
  \bibfield  {author} {\bibinfo {author} {\bibfnamefont {F.}~\bibnamefont
  {Nichele}}, \bibinfo {author} {\bibfnamefont {H.~J.}\ \bibnamefont
  {Suominen}}, \bibinfo {author} {\bibfnamefont {M.}~\bibnamefont
  {Kjaergaard}}, \bibinfo {author} {\bibfnamefont {C.~M.}\ \bibnamefont
  {Marcus}}, \bibinfo {author} {\bibfnamefont {E.}~\bibnamefont {Sajadi}},
  \bibinfo {author} {\bibfnamefont {J.~A.}\ \bibnamefont {Folk}}, \bibinfo
  {author} {\bibfnamefont {F.}~\bibnamefont {Qu}}, \bibinfo {author}
  {\bibfnamefont {A.~J.~A.}\ \bibnamefont {Beukman}}, \bibinfo {author}
  {\bibfnamefont {F.~K.~d.}\ \bibnamefont {Vries}}, \bibinfo {author}
  {\bibfnamefont {J.~v.}\ \bibnamefont {Veen}}, \bibinfo {author}
  {\bibfnamefont {S.}~\bibnamefont {Nadj-Perge}}, \bibinfo {author}
  {\bibfnamefont {L.~P.}\ \bibnamefont {Kouwenhoven}}, \bibinfo {author}
  {\bibfnamefont {B.-M.}\ \bibnamefont {Nguyen}}, \bibinfo {author}
  {\bibfnamefont {A.~A.}\ \bibnamefont {Kiselev}}, \bibinfo {author}
  {\bibfnamefont {W.}~\bibnamefont {Yi}}, \bibinfo {author} {\bibfnamefont
  {M.}~\bibnamefont {Sokolich}}, \bibinfo {author} {\bibfnamefont {M.~J.}\
  \bibnamefont {Manfra}}, \bibinfo {author} {\bibfnamefont {E.~M.}\
  \bibnamefont {Spanton}}, \ and\ \bibinfo {author} {\bibfnamefont {K.~A.}\
  \bibnamefont {Moler}},\ }\href {\doibase 10.1088/1367-2630/18/8/083005}
  {\bibfield  {journal} {\bibinfo  {journal} {New Journal of Physics}\ }\textbf
  {\bibinfo {volume} {18}},\ \bibinfo {pages} {083005} (\bibinfo {year}
  {2016})}\BibitemShut {NoStop}%
\bibitem [{\citenamefont {Nguyen}\ \emph {et~al.}(2016)\citenamefont {Nguyen},
  \citenamefont {Kiselev}, \citenamefont {Noah}, \citenamefont {Yi},
  \citenamefont {Qu}, \citenamefont {Beukman}, \citenamefont {de~Vries},
  \citenamefont {van Veen}, \citenamefont {Nadj-Perge}, \citenamefont
  {Kouwenhoven}, \citenamefont {Kjaergaard}, \citenamefont {Suominen},
  \citenamefont {Nichele}, \citenamefont {Marcus}, \citenamefont {Manfra},\
  and\ \citenamefont {Sokolich}}]{nguyen_decoupling_2016}%
  \BibitemOpen
  \bibfield  {author} {\bibinfo {author} {\bibfnamefont {B.-M.}\ \bibnamefont
  {Nguyen}}, \bibinfo {author} {\bibfnamefont {A.~A.}\ \bibnamefont {Kiselev}},
  \bibinfo {author} {\bibfnamefont {R.}~\bibnamefont {Noah}}, \bibinfo {author}
  {\bibfnamefont {W.}~\bibnamefont {Yi}}, \bibinfo {author} {\bibfnamefont
  {F.}~\bibnamefont {Qu}}, \bibinfo {author} {\bibfnamefont {A.~J.}\
  \bibnamefont {Beukman}}, \bibinfo {author} {\bibfnamefont {F.~K.}\
  \bibnamefont {de~Vries}}, \bibinfo {author} {\bibfnamefont {J.}~\bibnamefont
  {van Veen}}, \bibinfo {author} {\bibfnamefont {S.}~\bibnamefont
  {Nadj-Perge}}, \bibinfo {author} {\bibfnamefont {L.~P.}\ \bibnamefont
  {Kouwenhoven}}, \bibinfo {author} {\bibfnamefont {M.}~\bibnamefont
  {Kjaergaard}}, \bibinfo {author} {\bibfnamefont {H.~J.}\ \bibnamefont
  {Suominen}}, \bibinfo {author} {\bibfnamefont {F.}~\bibnamefont {Nichele}},
  \bibinfo {author} {\bibfnamefont {C.~M.}\ \bibnamefont {Marcus}}, \bibinfo
  {author} {\bibfnamefont {M.~J.}\ \bibnamefont {Manfra}}, \ and\ \bibinfo
  {author} {\bibfnamefont {M.}~\bibnamefont {Sokolich}},\ }\href {\doibase
  10.1103/PhysRevLett.117.077701} {\bibfield  {journal} {\bibinfo  {journal}
  {Physical Review Letters}\ }\textbf {\bibinfo {volume} {117}},\ \bibinfo
  {pages} {077701} (\bibinfo {year} {2016})}\BibitemShut {NoStop}%
\bibitem [{\citenamefont {Mueller}\ \emph {et~al.}(2017)\citenamefont
  {Mueller}, \citenamefont {Mittag}, \citenamefont {Tschirky}, \citenamefont
  {Charpentier}, \citenamefont {Wegscheider}, \citenamefont {Ensslin},\ and\
  \citenamefont {Ihn}}]{mueller_edge_2017}%
  \BibitemOpen
  \bibfield  {author} {\bibinfo {author} {\bibfnamefont {S.}~\bibnamefont
  {Mueller}}, \bibinfo {author} {\bibfnamefont {C.}~\bibnamefont {Mittag}},
  \bibinfo {author} {\bibfnamefont {T.}~\bibnamefont {Tschirky}}, \bibinfo
  {author} {\bibfnamefont {C.}~\bibnamefont {Charpentier}}, \bibinfo {author}
  {\bibfnamefont {W.}~\bibnamefont {Wegscheider}}, \bibinfo {author}
  {\bibfnamefont {K.}~\bibnamefont {Ensslin}}, \ and\ \bibinfo {author}
  {\bibfnamefont {T.}~\bibnamefont {Ihn}},\ }\href
  {http://arxiv.org/abs/1706.00320} {\bibfield  {journal} {\bibinfo  {journal}
  {arXiv:1706.00320 [cond-mat]}\ } (\bibinfo {year} {2017})},\ \bibinfo {note}
  {arXiv: 1706.00320}\BibitemShut {NoStop}%
\bibitem [{\citenamefont {Pfund}\ \emph {et~al.}(2006)\citenamefont {Pfund},
  \citenamefont {Shorubalko}, \citenamefont {Leturcq}, \citenamefont
  {Borgstr{\"o}m}, \citenamefont {Gramm}, \citenamefont {M{\"u}ller},\ and\
  \citenamefont {Ensslin}}]{pfund_fabrication_2006}%
  \BibitemOpen
  \bibfield  {author} {\bibinfo {author} {\bibfnamefont {A.}~\bibnamefont
  {Pfund}}, \bibinfo {author} {\bibfnamefont {I.}~\bibnamefont {Shorubalko}},
  \bibinfo {author} {\bibfnamefont {R.}~\bibnamefont {Leturcq}}, \bibinfo
  {author} {\bibfnamefont {M.~T.}\ \bibnamefont {Borgstr{\"o}m}}, \bibinfo
  {author} {\bibfnamefont {F.}~\bibnamefont {Gramm}}, \bibinfo {author}
  {\bibfnamefont {E.}~\bibnamefont {M{\"u}ller}}, \ and\ \bibinfo {author}
  {\bibfnamefont {K.}~\bibnamefont {Ensslin}},\ }\href {\doibase
  10.2533/chimia.2006.729} {\bibfield  {journal} {\bibinfo  {journal} {CHIMIA
  International Journal for Chemistry}\ }\textbf {\bibinfo {volume} {60}},\
  \bibinfo {pages} {729} (\bibinfo {year} {2006})}\BibitemShut {NoStop}%
\bibitem [{\citenamefont {Tilburg}\ \emph {et~al.}(2010)\citenamefont
  {Tilburg}, \citenamefont {Algra}, \citenamefont {Immink}, \citenamefont
  {Verheijen}, \citenamefont {Bakkers},\ and\ \citenamefont
  {Kouwenhoven}}]{tilburg_surface_2010}%
  \BibitemOpen
  \bibfield  {author} {\bibinfo {author} {\bibfnamefont {J.~W. W.~v.}\
  \bibnamefont {Tilburg}}, \bibinfo {author} {\bibfnamefont {R.~E.}\
  \bibnamefont {Algra}}, \bibinfo {author} {\bibfnamefont {W.~G.~G.}\
  \bibnamefont {Immink}}, \bibinfo {author} {\bibfnamefont {M.}~\bibnamefont
  {Verheijen}}, \bibinfo {author} {\bibfnamefont {E.~P. A.~M.}\ \bibnamefont
  {Bakkers}}, \ and\ \bibinfo {author} {\bibfnamefont {L.~P.}\ \bibnamefont
  {Kouwenhoven}},\ }\href {\doibase 10.1088/0268-1242/25/2/024011} {\bibfield
  {journal} {\bibinfo  {journal} {Semiconductor Science and Technology}\
  }\textbf {\bibinfo {volume} {25}},\ \bibinfo {pages} {024011} (\bibinfo
  {year} {2010})}\BibitemShut {NoStop}%
\bibitem [{\citenamefont {Punkkinen}\ \emph {et~al.}(2011)\citenamefont
  {Punkkinen}, \citenamefont {Laukkanen}, \citenamefont {L{\aa}ng},
  \citenamefont {Kuzmin}, \citenamefont {Tuominen}, \citenamefont {Tuominen},
  \citenamefont {Dahl}, \citenamefont {Pessa}, \citenamefont {Guina},
  \citenamefont {Kokko}, \citenamefont {Sadowski}, \citenamefont {Johansson},
  \citenamefont {V{\"a}yrynen},\ and\ \citenamefont
  {Vitos}}]{punkkinen_oxidized_2011}%
  \BibitemOpen
  \bibfield  {author} {\bibinfo {author} {\bibfnamefont {M.~P.~J.}\
  \bibnamefont {Punkkinen}}, \bibinfo {author} {\bibfnamefont {P.}~\bibnamefont
  {Laukkanen}}, \bibinfo {author} {\bibfnamefont {J.}~\bibnamefont {L{\aa}ng}},
  \bibinfo {author} {\bibfnamefont {M.}~\bibnamefont {Kuzmin}}, \bibinfo
  {author} {\bibfnamefont {M.}~\bibnamefont {Tuominen}}, \bibinfo {author}
  {\bibfnamefont {V.}~\bibnamefont {Tuominen}}, \bibinfo {author}
  {\bibfnamefont {J.}~\bibnamefont {Dahl}}, \bibinfo {author} {\bibfnamefont
  {M.}~\bibnamefont {Pessa}}, \bibinfo {author} {\bibfnamefont
  {M.}~\bibnamefont {Guina}}, \bibinfo {author} {\bibfnamefont
  {K.}~\bibnamefont {Kokko}}, \bibinfo {author} {\bibfnamefont
  {J.}~\bibnamefont {Sadowski}}, \bibinfo {author} {\bibfnamefont
  {B.}~\bibnamefont {Johansson}}, \bibinfo {author} {\bibfnamefont {I.~J.}\
  \bibnamefont {V{\"a}yrynen}}, \ and\ \bibinfo {author} {\bibfnamefont
  {L.}~\bibnamefont {Vitos}},\ }\href {\doibase 10.1103/PhysRevB.83.195329}
  {\bibfield  {journal} {\bibinfo  {journal} {Physical Review B}\ }\textbf
  {\bibinfo {volume} {83}},\ \bibinfo {pages} {195329} (\bibinfo {year}
  {2011})}\BibitemShut {NoStop}%
\bibitem [{\citenamefont {Wang}\ \emph {et~al.}(2013)\citenamefont {Wang},
  \citenamefont {Wang}, \citenamefont {Doornbos}, \citenamefont {Astromskas},
  \citenamefont {Bhuwalka}, \citenamefont {Contreras-Guerrero}, \citenamefont
  {Edirisooriya}, \citenamefont {Rojas-Ramirez}, \citenamefont {Vellianitis},
  \citenamefont {Oxland}, \citenamefont {Holland}, \citenamefont {Hsieh},
  \citenamefont {Ramvall}, \citenamefont {Lind}, \citenamefont {Hsu},
  \citenamefont {Wernersson}, \citenamefont {Droopad}, \citenamefont
  {Passlack},\ and\ \citenamefont {Diaz}}]{wang_inas_2013}%
  \BibitemOpen
  \bibfield  {author} {\bibinfo {author} {\bibfnamefont {C.~H.}\ \bibnamefont
  {Wang}}, \bibinfo {author} {\bibfnamefont {S.~W.}\ \bibnamefont {Wang}},
  \bibinfo {author} {\bibfnamefont {G.}~\bibnamefont {Doornbos}}, \bibinfo
  {author} {\bibfnamefont {G.}~\bibnamefont {Astromskas}}, \bibinfo {author}
  {\bibfnamefont {K.}~\bibnamefont {Bhuwalka}}, \bibinfo {author}
  {\bibfnamefont {R.}~\bibnamefont {Contreras-Guerrero}}, \bibinfo {author}
  {\bibfnamefont {M.}~\bibnamefont {Edirisooriya}}, \bibinfo {author}
  {\bibfnamefont {J.~S.}\ \bibnamefont {Rojas-Ramirez}}, \bibinfo {author}
  {\bibfnamefont {G.}~\bibnamefont {Vellianitis}}, \bibinfo {author}
  {\bibfnamefont {R.}~\bibnamefont {Oxland}}, \bibinfo {author} {\bibfnamefont
  {M.~C.}\ \bibnamefont {Holland}}, \bibinfo {author} {\bibfnamefont {C.~H.}\
  \bibnamefont {Hsieh}}, \bibinfo {author} {\bibfnamefont {P.}~\bibnamefont
  {Ramvall}}, \bibinfo {author} {\bibfnamefont {E.}~\bibnamefont {Lind}},
  \bibinfo {author} {\bibfnamefont {W.~C.}\ \bibnamefont {Hsu}}, \bibinfo
  {author} {\bibfnamefont {L.-E.}\ \bibnamefont {Wernersson}}, \bibinfo
  {author} {\bibfnamefont {R.}~\bibnamefont {Droopad}}, \bibinfo {author}
  {\bibfnamefont {M.}~\bibnamefont {Passlack}}, \ and\ \bibinfo {author}
  {\bibfnamefont {C.~H.}\ \bibnamefont {Diaz}},\ }\href {\doibase
  10.1063/1.4820477} {\bibfield  {journal} {\bibinfo  {journal} {Applied
  Physics Letters}\ }\textbf {\bibinfo {volume} {103}},\ \bibinfo {pages}
  {143510} (\bibinfo {year} {2013})}\BibitemShut {NoStop}%
\bibitem [{\citenamefont {Oxland}\ \emph {et~al.}(2016)\citenamefont {Oxland},
  \citenamefont {Li}, \citenamefont {Chang}, \citenamefont {Wang},
  \citenamefont {Vasen}, \citenamefont {Ramvall}, \citenamefont
  {Contreras-Guerrero}, \citenamefont {Rojas-Ramirez}, \citenamefont {Holland},
  \citenamefont {Doornbos}, \citenamefont {Chang}, \citenamefont {Macintyre},
  \citenamefont {Thoms}, \citenamefont {Droopad}, \citenamefont {Yeo},
  \citenamefont {Diaz}, \citenamefont {Thayne},\ and\ \citenamefont
  {Passlack}}]{oxland_inas_2016}%
  \BibitemOpen
  \bibfield  {author} {\bibinfo {author} {\bibfnamefont {R.}~\bibnamefont
  {Oxland}}, \bibinfo {author} {\bibfnamefont {X.}~\bibnamefont {Li}}, \bibinfo
  {author} {\bibfnamefont {S.~W.}\ \bibnamefont {Chang}}, \bibinfo {author}
  {\bibfnamefont {S.~W.}\ \bibnamefont {Wang}}, \bibinfo {author}
  {\bibfnamefont {T.}~\bibnamefont {Vasen}}, \bibinfo {author} {\bibfnamefont
  {P.}~\bibnamefont {Ramvall}}, \bibinfo {author} {\bibfnamefont
  {R.}~\bibnamefont {Contreras-Guerrero}}, \bibinfo {author} {\bibfnamefont
  {J.}~\bibnamefont {Rojas-Ramirez}}, \bibinfo {author} {\bibfnamefont
  {M.}~\bibnamefont {Holland}}, \bibinfo {author} {\bibfnamefont
  {G.}~\bibnamefont {Doornbos}}, \bibinfo {author} {\bibfnamefont {Y.~S.}\
  \bibnamefont {Chang}}, \bibinfo {author} {\bibfnamefont {D.~S.}\ \bibnamefont
  {Macintyre}}, \bibinfo {author} {\bibfnamefont {S.}~\bibnamefont {Thoms}},
  \bibinfo {author} {\bibfnamefont {R.}~\bibnamefont {Droopad}}, \bibinfo
  {author} {\bibfnamefont {Y.~C.}\ \bibnamefont {Yeo}}, \bibinfo {author}
  {\bibfnamefont {C.~H.}\ \bibnamefont {Diaz}}, \bibinfo {author}
  {\bibfnamefont {I.~G.}\ \bibnamefont {Thayne}}, \ and\ \bibinfo {author}
  {\bibfnamefont {M.}~\bibnamefont {Passlack}},\ }\href {\doibase
  10.1109/LED.2016.2521001} {\bibfield  {journal} {\bibinfo  {journal} {IEEE
  Electron Device Letters}\ }\textbf {\bibinfo {volume} {37}},\ \bibinfo
  {pages} {261} (\bibinfo {year} {2016})}\BibitemShut {NoStop}%
\bibitem [{\citenamefont {Chaghi}\ \emph {et~al.}(2009)\citenamefont {Chaghi},
  \citenamefont {Cervera}, \citenamefont {Aït-Kaci}, \citenamefont {Grech},
  \citenamefont {Rodriguez},\ and\ \citenamefont {Christol}}]{chaghi_wet_2009}%
  \BibitemOpen
  \bibfield  {author} {\bibinfo {author} {\bibfnamefont {R.}~\bibnamefont
  {Chaghi}}, \bibinfo {author} {\bibfnamefont {C.}~\bibnamefont {Cervera}},
  \bibinfo {author} {\bibfnamefont {H.}~\bibnamefont {Aït-Kaci}}, \bibinfo
  {author} {\bibfnamefont {P.}~\bibnamefont {Grech}}, \bibinfo {author}
  {\bibfnamefont {J.~B.}\ \bibnamefont {Rodriguez}}, \ and\ \bibinfo {author}
  {\bibfnamefont {P.}~\bibnamefont {Christol}},\ }\href {\doibase
  10.1088/0268-1242/24/6/065010} {\bibfield  {journal} {\bibinfo  {journal}
  {Semiconductor Science and Technology}\ }\textbf {\bibinfo {volume} {24}},\
  \bibinfo {pages} {065010} (\bibinfo {year} {2009})}\BibitemShut {NoStop}%
\bibitem [{\citenamefont {Pal}\ \emph {et~al.}(2015)\citenamefont {Pal},
  \citenamefont {Mueller}, \citenamefont {Ihn}, \citenamefont {Ensslin},
  \citenamefont {Tschirky}, \citenamefont {Charpentier},\ and\ \citenamefont
  {Wegscheider}}]{pal_influence_2015}%
  \BibitemOpen
  \bibfield  {author} {\bibinfo {author} {\bibfnamefont {A.~N.}\ \bibnamefont
  {Pal}}, \bibinfo {author} {\bibfnamefont {S.}~\bibnamefont {Mueller}},
  \bibinfo {author} {\bibfnamefont {T.}~\bibnamefont {Ihn}}, \bibinfo {author}
  {\bibfnamefont {K.}~\bibnamefont {Ensslin}}, \bibinfo {author} {\bibfnamefont
  {T.}~\bibnamefont {Tschirky}}, \bibinfo {author} {\bibfnamefont
  {C.}~\bibnamefont {Charpentier}}, \ and\ \bibinfo {author} {\bibfnamefont
  {W.}~\bibnamefont {Wegscheider}},\ }\href {\doibase 10.1063/1.4926385}
  {\bibfield  {journal} {\bibinfo  {journal} {AIP Advances}\ }\textbf {\bibinfo
  {volume} {5}},\ \bibinfo {pages} {077106} (\bibinfo {year}
  {2015})}\BibitemShut {NoStop}%
\bibitem [{\citenamefont {Plis}, \citenamefont {Kutty},\ and\ \citenamefont
  {Krishna}(2013)}]{plis_passivation_2013}%
  \BibitemOpen
  \bibfield  {author} {\bibinfo {author} {\bibfnamefont {E.~A.}\ \bibnamefont
  {Plis}}, \bibinfo {author} {\bibfnamefont {M.~N.}\ \bibnamefont {Kutty}}, \
  and\ \bibinfo {author} {\bibfnamefont {S.}~\bibnamefont {Krishna}},\ }\href
  {\doibase 10.1002/lpor.201100029} {\bibfield  {journal} {\bibinfo  {journal}
  {Laser \& Photonics Reviews}\ }\textbf {\bibinfo {volume} {7}},\ \bibinfo
  {pages} {45} (\bibinfo {year} {2013})}\BibitemShut {NoStop}%
\bibitem [{\citenamefont {Petrovykh}, \citenamefont {Yang},\ and\ \citenamefont
  {Whitman}(2003)}]{petrovykh_chemical_2003}%
  \BibitemOpen
  \bibfield  {author} {\bibinfo {author} {\bibfnamefont {D.~Y.}\ \bibnamefont
  {Petrovykh}}, \bibinfo {author} {\bibfnamefont {M.~J.}\ \bibnamefont {Yang}},
  \ and\ \bibinfo {author} {\bibfnamefont {L.~J.}\ \bibnamefont {Whitman}},\
  }\href {\doibase 10.1016/S0039-6028(02)02411-1} {\bibfield  {journal}
  {\bibinfo  {journal} {Surface Science}\ }\textbf {\bibinfo {volume} {523}},\
  \bibinfo {pages} {231} (\bibinfo {year} {2003})}\BibitemShut {NoStop}%
\bibitem [{\citenamefont {Petrovykh}, \citenamefont {Long},\ and\ \citenamefont
  {Whitman}(2005)}]{petrovykh_surface_2005}%
  \BibitemOpen
  \bibfield  {author} {\bibinfo {author} {\bibfnamefont {D.~Y.}\ \bibnamefont
  {Petrovykh}}, \bibinfo {author} {\bibfnamefont {J.~P.}\ \bibnamefont {Long}},
  \ and\ \bibinfo {author} {\bibfnamefont {L.~J.}\ \bibnamefont {Whitman}},\
  }\href {\doibase 10.1063/1.1946182} {\bibfield  {journal} {\bibinfo
  {journal} {Applied Physics Letters}\ }\textbf {\bibinfo {volume} {86}},\
  \bibinfo {pages} {242105} (\bibinfo {year} {2005})}\BibitemShut {NoStop}%
\bibitem [{\citenamefont {Hang}\ \emph {et~al.}(2008)\citenamefont {Hang},
  \citenamefont {Wang}, \citenamefont {Carpenter}, \citenamefont {Zemlyanov},
  \citenamefont {Zakharov}, \citenamefont {Stach}, \citenamefont {Buhro},\ and\
  \citenamefont {Janes}}]{hang_role_2008}%
  \BibitemOpen
  \bibfield  {author} {\bibinfo {author} {\bibfnamefont {Q.}~\bibnamefont
  {Hang}}, \bibinfo {author} {\bibfnamefont {F.}~\bibnamefont {Wang}}, \bibinfo
  {author} {\bibfnamefont {P.~D.}\ \bibnamefont {Carpenter}}, \bibinfo {author}
  {\bibfnamefont {D.}~\bibnamefont {Zemlyanov}}, \bibinfo {author}
  {\bibfnamefont {D.}~\bibnamefont {Zakharov}}, \bibinfo {author}
  {\bibfnamefont {E.~A.}\ \bibnamefont {Stach}}, \bibinfo {author}
  {\bibfnamefont {W.~E.}\ \bibnamefont {Buhro}}, \ and\ \bibinfo {author}
  {\bibfnamefont {D.~B.}\ \bibnamefont {Janes}},\ }\href {\doibase
  10.1021/nl071888t} {\bibfield  {journal} {\bibinfo  {journal} {Nano Letters}\
  }\textbf {\bibinfo {volume} {8}},\ \bibinfo {pages} {49} (\bibinfo {year}
  {2008})}\BibitemShut {NoStop}%
\bibitem [{\citenamefont {Sun}\ \emph {et~al.}(2012)\citenamefont {Sun},
  \citenamefont {Joyce}, \citenamefont {Gao}, \citenamefont {Tan},
  \citenamefont {Jagadish},\ and\ \citenamefont {Ning}}]{sun_removal_2012}%
  \BibitemOpen
  \bibfield  {author} {\bibinfo {author} {\bibfnamefont {M.~H.}\ \bibnamefont
  {Sun}}, \bibinfo {author} {\bibfnamefont {H.~J.}\ \bibnamefont {Joyce}},
  \bibinfo {author} {\bibfnamefont {Q.}~\bibnamefont {Gao}}, \bibinfo {author}
  {\bibfnamefont {H.~H.}\ \bibnamefont {Tan}}, \bibinfo {author} {\bibfnamefont
  {C.}~\bibnamefont {Jagadish}}, \ and\ \bibinfo {author} {\bibfnamefont
  {C.~Z.}\ \bibnamefont {Ning}},\ }\href {\doibase 10.1021/nl300015w}
  {\bibfield  {journal} {\bibinfo  {journal} {Nano Letters}\ }\textbf {\bibinfo
  {volume} {12}},\ \bibinfo {pages} {3378} (\bibinfo {year}
  {2012})}\BibitemShut {NoStop}%
\bibitem [{\citenamefont {Pi}\ \emph {et~al.}(1999)\citenamefont {Pi},
  \citenamefont {Lin}, \citenamefont {Chang}, \citenamefont {Chang},
  \citenamefont {Hong}, \citenamefont {Kwo},\ and\ \citenamefont
  {Webster}}]{pi_semiconductor-insulator_1999}%
  \BibitemOpen
  \bibfield  {author} {\bibinfo {author} {\bibfnamefont {T.~W.}\ \bibnamefont
  {Pi}}, \bibinfo {author} {\bibfnamefont {T.~D.}\ \bibnamefont {Lin}},
  \bibinfo {author} {\bibfnamefont {W.~H.}\ \bibnamefont {Chang}}, \bibinfo
  {author} {\bibfnamefont {Y.~C.}\ \bibnamefont {Chang}}, \bibinfo {author}
  {\bibfnamefont {M.}~\bibnamefont {Hong}}, \bibinfo {author} {\bibfnamefont
  {J.}~\bibnamefont {Kwo}}, \ and\ \bibinfo {author} {\bibfnamefont {J.~G.}\
  \bibnamefont {Webster}},\ }in\ \href
  {http://onlinelibrary.wiley.com/doi/10.1002/047134608X.W3226.pub2/abstract}
  {\emph {\bibinfo {booktitle} {Wiley {Encyclopedia} of {Electrical} and
  {Electronics} {Engineering}}}}\ (\bibinfo  {publisher} {John Wiley \& Sons,
  Inc.},\ \bibinfo {year} {1999})\ \bibinfo {note} {dOI:
  10.1002/047134608X.W3226.pub2}\BibitemShut {NoStop}%
\bibitem [{\citenamefont {Lee}\ \emph {et~al.}(2010)\citenamefont {Lee},
  \citenamefont {Feng}, \citenamefont {Yu}, \citenamefont {Mastrogiovanni},
  \citenamefont {Wan}, \citenamefont {Garfunkel},\ and\ \citenamefont
  {Gustafsson}}]{lee_ald_2010}%
  \BibitemOpen
  \bibfield  {author} {\bibinfo {author} {\bibfnamefont {H.~D.}\ \bibnamefont
  {Lee}}, \bibinfo {author} {\bibfnamefont {T.}~\bibnamefont {Feng}}, \bibinfo
  {author} {\bibfnamefont {L.}~\bibnamefont {Yu}}, \bibinfo {author}
  {\bibfnamefont {D.}~\bibnamefont {Mastrogiovanni}}, \bibinfo {author}
  {\bibfnamefont {A.}~\bibnamefont {Wan}}, \bibinfo {author} {\bibfnamefont
  {E.}~\bibnamefont {Garfunkel}}, \ and\ \bibinfo {author} {\bibfnamefont
  {T.}~\bibnamefont {Gustafsson}},\ }\href {\doibase 10.1002/pssc.200982425}
  {\bibfield  {journal} {\bibinfo  {journal} {physica status solidi (c)}\
  }\textbf {\bibinfo {volume} {7}},\ \bibinfo {pages} {260} (\bibinfo {year}
  {2010})}\BibitemShut {NoStop}%
\bibitem [{\citenamefont {Tallarida}\ \emph {et~al.}(2011)\citenamefont
  {Tallarida}, \citenamefont {Adelmann}, \citenamefont {Delabie}, \citenamefont
  {Van~Elshocht}, \citenamefont {Caymax},\ and\ \citenamefont
  {Schmeisser}}]{tallarida_surface_2011}%
  \BibitemOpen
  \bibfield  {author} {\bibinfo {author} {\bibfnamefont {M.}~\bibnamefont
  {Tallarida}}, \bibinfo {author} {\bibfnamefont {C.}~\bibnamefont {Adelmann}},
  \bibinfo {author} {\bibfnamefont {A.}~\bibnamefont {Delabie}}, \bibinfo
  {author} {\bibfnamefont {S.}~\bibnamefont {Van~Elshocht}}, \bibinfo {author}
  {\bibfnamefont {M.}~\bibnamefont {Caymax}}, \ and\ \bibinfo {author}
  {\bibfnamefont {D.}~\bibnamefont {Schmeisser}},\ }\href {\doibase
  10.1063/1.3615784} {\bibfield  {journal} {\bibinfo  {journal} {Applied
  Physics Letters}\ }\textbf {\bibinfo {volume} {99}},\ \bibinfo {pages}
  {042906} (\bibinfo {year} {2011})}\BibitemShut {NoStop}%
\bibitem [{\citenamefont {Konda}\ \emph {et~al.}(2011)\citenamefont {Konda},
  \citenamefont {Mundle}, \citenamefont {Bamiduro}, \citenamefont {Dondapati},
  \citenamefont {Bahoura}, \citenamefont {Pradhan},\ and\ \citenamefont
  {Donley}}]{konda_effect_2011}%
  \BibitemOpen
  \bibfield  {author} {\bibinfo {author} {\bibfnamefont {R.~B.}\ \bibnamefont
  {Konda}}, \bibinfo {author} {\bibfnamefont {R.}~\bibnamefont {Mundle}},
  \bibinfo {author} {\bibfnamefont {O.}~\bibnamefont {Bamiduro}}, \bibinfo
  {author} {\bibfnamefont {H.}~\bibnamefont {Dondapati}}, \bibinfo {author}
  {\bibfnamefont {M.}~\bibnamefont {Bahoura}}, \bibinfo {author} {\bibfnamefont
  {A.~K.}\ \bibnamefont {Pradhan}}, \ and\ \bibinfo {author} {\bibfnamefont
  {C.}~\bibnamefont {Donley}},\ }\href {\doibase 10.1116/1.3662862} {\bibfield
  {journal} {\bibinfo  {journal} {Journal of Vacuum Science \& Technology A:
  Vacuum, Surfaces, and Films}\ }\textbf {\bibinfo {volume} {30}},\ \bibinfo
  {pages} {01A118} (\bibinfo {year} {2011})}\BibitemShut {NoStop}%
\bibitem [{\citenamefont {Trinh}\ \emph {et~al.}(2011)\citenamefont {Trinh},
  \citenamefont {Chang}, \citenamefont {Brammertz}, \citenamefont {Lu},
  \citenamefont {Nguyen},\ and\ \citenamefont
  {Tran}}]{trinh_experimental_2011}%
  \BibitemOpen
  \bibfield  {author} {\bibinfo {author} {\bibfnamefont {H.~D.}\ \bibnamefont
  {Trinh}}, \bibinfo {author} {\bibfnamefont {E.~Y.}\ \bibnamefont {Chang}},
  \bibinfo {author} {\bibfnamefont {G.}~\bibnamefont {Brammertz}}, \bibinfo
  {author} {\bibfnamefont {C.-Y.}\ \bibnamefont {Lu}}, \bibinfo {author}
  {\bibfnamefont {H.-Q.}\ \bibnamefont {Nguyen}}, \ and\ \bibinfo {author}
  {\bibfnamefont {B.-T.}\ \bibnamefont {Tran}},\ }\href {\doibase
  10.1149/1.3567712} {\bibfield  {journal} {\bibinfo  {journal} {ECS
  Transactions}\ }\textbf {\bibinfo {volume} {34}},\ \bibinfo {pages} {1041}
  (\bibinfo {year} {2011})}\BibitemShut {NoStop}%
\bibitem [{\citenamefont {Hollinger}, \citenamefont {Skheyta-Kabbani},\ and\
  \citenamefont {Gendry}(1994)}]{hollinger_oxides_1994}%
  \BibitemOpen
  \bibfield  {author} {\bibinfo {author} {\bibfnamefont {G.}~\bibnamefont
  {Hollinger}}, \bibinfo {author} {\bibfnamefont {R.}~\bibnamefont
  {Skheyta-Kabbani}}, \ and\ \bibinfo {author} {\bibfnamefont {M.}~\bibnamefont
  {Gendry}},\ }\href {\doibase 10.1103/PhysRevB.49.11159} {\bibfield  {journal}
  {\bibinfo  {journal} {Physical Review B}\ }\textbf {\bibinfo {volume} {49}},\
  \bibinfo {pages} {11159} (\bibinfo {year} {1994})}\BibitemShut {NoStop}%
\bibitem [{\citenamefont {Timm}\ \emph {et~al.}(2010)\citenamefont {Timm},
  \citenamefont {Fian}, \citenamefont {Hjort}, \citenamefont {Thelander},
  \citenamefont {Lind}, \citenamefont {Andersen}, \citenamefont {Wernersson},\
  and\ \citenamefont {Mikkelsen}}]{timm_reduction_2010}%
  \BibitemOpen
  \bibfield  {author} {\bibinfo {author} {\bibfnamefont {R.}~\bibnamefont
  {Timm}}, \bibinfo {author} {\bibfnamefont {A.}~\bibnamefont {Fian}}, \bibinfo
  {author} {\bibfnamefont {M.}~\bibnamefont {Hjort}}, \bibinfo {author}
  {\bibfnamefont {C.}~\bibnamefont {Thelander}}, \bibinfo {author}
  {\bibfnamefont {E.}~\bibnamefont {Lind}}, \bibinfo {author} {\bibfnamefont
  {J.~N.}\ \bibnamefont {Andersen}}, \bibinfo {author} {\bibfnamefont {L.-E.}\
  \bibnamefont {Wernersson}}, \ and\ \bibinfo {author} {\bibfnamefont
  {A.}~\bibnamefont {Mikkelsen}},\ }\href {\doibase 10.1063/1.3495776}
  {\bibfield  {journal} {\bibinfo  {journal} {Applied Physics Letters}\
  }\textbf {\bibinfo {volume} {97}},\ \bibinfo {pages} {132904} (\bibinfo
  {year} {2010})}\BibitemShut {NoStop}%
\bibitem [{\citenamefont {Salihoglu}\ \emph {et~al.}(2012)\citenamefont
  {Salihoglu}, \citenamefont {Muti}, \citenamefont {Kutluer}, \citenamefont
  {Tansel}, \citenamefont {Turan}, \citenamefont {Kocabas},\ and\ \citenamefont
  {Aydinli}}]{salihoglu_atomic_2012}%
  \BibitemOpen
  \bibfield  {author} {\bibinfo {author} {\bibfnamefont {O.}~\bibnamefont
  {Salihoglu}}, \bibinfo {author} {\bibfnamefont {A.}~\bibnamefont {Muti}},
  \bibinfo {author} {\bibfnamefont {K.}~\bibnamefont {Kutluer}}, \bibinfo
  {author} {\bibfnamefont {T.}~\bibnamefont {Tansel}}, \bibinfo {author}
  {\bibfnamefont {R.}~\bibnamefont {Turan}}, \bibinfo {author} {\bibfnamefont
  {C.}~\bibnamefont {Kocabas}}, \ and\ \bibinfo {author} {\bibfnamefont
  {A.}~\bibnamefont {Aydinli}},\ }\href {\doibase 10.1063/1.3702567} {\bibfield
   {journal} {\bibinfo  {journal} {Journal of Applied Physics}\ }\textbf
  {\bibinfo {volume} {111}},\ \bibinfo {pages} {074509} (\bibinfo {year}
  {2012})}\BibitemShut {NoStop}%
\bibitem [{\citenamefont {Trinh}\ \emph {et~al.}(2010)\citenamefont {Trinh},
  \citenamefont {Chang}, \citenamefont {Wong}, \citenamefont {Yu},
  \citenamefont {Chang}, \citenamefont {Lin}, \citenamefont {Nguyen},\ and\
  \citenamefont {Tran}}]{trinh_effects_2010}%
  \BibitemOpen
  \bibfield  {author} {\bibinfo {author} {\bibfnamefont {H.-D.}\ \bibnamefont
  {Trinh}}, \bibinfo {author} {\bibfnamefont {E.~Y.}\ \bibnamefont {Chang}},
  \bibinfo {author} {\bibfnamefont {Y.-Y.}\ \bibnamefont {Wong}}, \bibinfo
  {author} {\bibfnamefont {C.-C.}\ \bibnamefont {Yu}}, \bibinfo {author}
  {\bibfnamefont {C.-Y.}\ \bibnamefont {Chang}}, \bibinfo {author}
  {\bibfnamefont {Y.-C.}\ \bibnamefont {Lin}}, \bibinfo {author} {\bibfnamefont
  {H.-Q.}\ \bibnamefont {Nguyen}}, \ and\ \bibinfo {author} {\bibfnamefont
  {B.-T.}\ \bibnamefont {Tran}},\ }\href {\doibase 10.1143/JJAP.49.111201}
  {\bibfield  {journal} {\bibinfo  {journal} {Japanese Journal of Applied
  Physics}\ }\textbf {\bibinfo {volume} {49}},\ \bibinfo {pages} {111201}
  (\bibinfo {year} {2010})}\BibitemShut {NoStop}%
\bibitem [{\citenamefont {Pulver}\ \emph {et~al.}(2001)\citenamefont {Pulver},
  \citenamefont {Wilmsen}, \citenamefont {Niles},\ and\ \citenamefont
  {Kee}}]{pulver_thermal_2001}%
  \BibitemOpen
  \bibfield  {author} {\bibinfo {author} {\bibfnamefont {D.}~\bibnamefont
  {Pulver}}, \bibinfo {author} {\bibfnamefont {C.~W.}\ \bibnamefont {Wilmsen}},
  \bibinfo {author} {\bibfnamefont {D.}~\bibnamefont {Niles}}, \ and\ \bibinfo
  {author} {\bibfnamefont {R.}~\bibnamefont {Kee}},\ }\href {\doibase
  10.1116/1.1342008} {\bibfield  {journal} {\bibinfo  {journal} {Journal of
  Vacuum Science \& Technology B: Microelectronics and Nanometer Structures
  Processing, Measurement, and Phenomena}\ }\textbf {\bibinfo {volume} {19}},\
  \bibinfo {pages} {207} (\bibinfo {year} {2001})}\BibitemShut {NoStop}%
\bibitem [{\citenamefont {Juppo}\ \emph {et~al.}(2001)\citenamefont {Juppo},
  \citenamefont {Al{\'e}n}, \citenamefont {Ritala},\ and\ \citenamefont
  {Leskel{\"a}}}]{juppo_trimethylaluminum_2001}%
  \BibitemOpen
  \bibfield  {author} {\bibinfo {author} {\bibfnamefont {M.}~\bibnamefont
  {Juppo}}, \bibinfo {author} {\bibfnamefont {P.}~\bibnamefont {Al{\'e}n}},
  \bibinfo {author} {\bibfnamefont {M.}~\bibnamefont {Ritala}}, \ and\ \bibinfo
  {author} {\bibfnamefont {M.}~\bibnamefont {Leskel{\"a}}},\ }\href {\doibase
  10.1002/1521-3862(200109)7:5<211::AID-CVDE211>3.0.CO;2-L} {\bibfield
  {journal} {\bibinfo  {journal} {Chemical Vapor Deposition}\ }\textbf
  {\bibinfo {volume} {7}},\ \bibinfo {pages} {211} (\bibinfo {year}
  {2001})}\BibitemShut {NoStop}%
\bibitem [{\citenamefont {Al{\'e}n}\ \emph {et~al.}(2001)\citenamefont
  {Al{\'e}n}, \citenamefont {Juppo}, \citenamefont {Ritala}, \citenamefont
  {Sajavaara}, \citenamefont {Keinonen},\ and\ \citenamefont
  {Leskel{\"a}}}]{alen_atomic_2001}%
  \BibitemOpen
  \bibfield  {author} {\bibinfo {author} {\bibfnamefont {P.}~\bibnamefont
  {Al{\'e}n}}, \bibinfo {author} {\bibfnamefont {M.}~\bibnamefont {Juppo}},
  \bibinfo {author} {\bibfnamefont {M.}~\bibnamefont {Ritala}}, \bibinfo
  {author} {\bibfnamefont {T.}~\bibnamefont {Sajavaara}}, \bibinfo {author}
  {\bibfnamefont {J.}~\bibnamefont {Keinonen}}, \ and\ \bibinfo {author}
  {\bibfnamefont {M.}~\bibnamefont {Leskel{\"a}}},\ }\href {\doibase
  10.1149/1.1401082} {\bibfield  {journal} {\bibinfo  {journal} {Journal of The
  Electrochemical Society}\ }\textbf {\bibinfo {volume} {148}},\ \bibinfo
  {pages} {G566} (\bibinfo {year} {2001})}\BibitemShut {NoStop}%
\bibitem [{\citenamefont {Lowe}\ \emph {et~al.}(2002)\citenamefont {Lowe},
  \citenamefont {Veal}, \citenamefont {McConville}, \citenamefont {Bell},
  \citenamefont {Tsukamoto},\ and\ \citenamefont
  {Koguchi}}]{lowe_extreme_2002}%
  \BibitemOpen
  \bibfield  {author} {\bibinfo {author} {\bibfnamefont {M.~J.}\ \bibnamefont
  {Lowe}}, \bibinfo {author} {\bibfnamefont {T.~D.}\ \bibnamefont {Veal}},
  \bibinfo {author} {\bibfnamefont {C.~F.}\ \bibnamefont {McConville}},
  \bibinfo {author} {\bibfnamefont {G.~R.}\ \bibnamefont {Bell}}, \bibinfo
  {author} {\bibfnamefont {S.}~\bibnamefont {Tsukamoto}}, \ and\ \bibinfo
  {author} {\bibfnamefont {N.}~\bibnamefont {Koguchi}},\ }\href {\doibase
  10.1016/S0022-0248(01)01899-1} {\bibfield  {journal} {\bibinfo  {journal}
  {Journal of Crystal Growth}\ }\bibinfo {series} {The thirteenth international
  conference on {Crystal} {Growth} in conj unction with the eleventh
  international conference on {Vapor} {Growth} and {Epitaxy}},\ \textbf
  {\bibinfo {volume} {237–239, Part 1}},\ \bibinfo {pages} {196} (\bibinfo
  {year} {2002})}\BibitemShut {NoStop}%
\bibitem [{\citenamefont {Lowe}\ \emph {et~al.}(2003)\citenamefont {Lowe},
  \citenamefont {Veal}, \citenamefont {Mowbray},\ and\ \citenamefont
  {McConville}}]{lowe_sulphur-induced_2003}%
  \BibitemOpen
  \bibfield  {author} {\bibinfo {author} {\bibfnamefont {M.~J.}\ \bibnamefont
  {Lowe}}, \bibinfo {author} {\bibfnamefont {T.~D.}\ \bibnamefont {Veal}},
  \bibinfo {author} {\bibfnamefont {A.~P.}\ \bibnamefont {Mowbray}}, \ and\
  \bibinfo {author} {\bibfnamefont {C.~F.}\ \bibnamefont {McConville}},\ }\href
  {\doibase 10.1016/j.susc.2003.08.047} {\bibfield  {journal} {\bibinfo
  {journal} {Surface Science}\ }\textbf {\bibinfo {volume} {544}},\ \bibinfo
  {pages} {320} (\bibinfo {year} {2003})}\BibitemShut {NoStop}%
\bibitem [{\citenamefont {Watanabe}\ and\ \citenamefont
  {Maeda}(1997)}]{watanabe_anomalous_1997}%
  \BibitemOpen
  \bibfield  {author} {\bibinfo {author} {\bibfnamefont {Y.}~\bibnamefont
  {Watanabe}}\ and\ \bibinfo {author} {\bibfnamefont {F.}~\bibnamefont
  {Maeda}},\ }\href {\doibase 10.1016/S0169-4332(97)80174-2} {\bibfield
  {journal} {\bibinfo  {journal} {Applied Surface Science}\ }\textbf {\bibinfo
  {volume} {117}},\ \bibinfo {pages} {735} (\bibinfo {year}
  {1997})}\BibitemShut {NoStop}%
\bibitem [{\citenamefont {Rehm}\ \emph {et~al.}(2005)\citenamefont {Rehm},
  \citenamefont {Walther}, \citenamefont {Fuchs}, \citenamefont {Schmitz},\
  and\ \citenamefont {Fleissner}}]{rehm_passivation_2005}%
  \BibitemOpen
  \bibfield  {author} {\bibinfo {author} {\bibfnamefont {R.}~\bibnamefont
  {Rehm}}, \bibinfo {author} {\bibfnamefont {M.}~\bibnamefont {Walther}},
  \bibinfo {author} {\bibfnamefont {F.}~\bibnamefont {Fuchs}}, \bibinfo
  {author} {\bibfnamefont {J.}~\bibnamefont {Schmitz}}, \ and\ \bibinfo
  {author} {\bibfnamefont {J.}~\bibnamefont {Fleissner}},\ }\href {\doibase
  10.1063/1.1906326} {\bibfield  {journal} {\bibinfo  {journal} {Applied
  Physics Letters}\ }\textbf {\bibinfo {volume} {86}},\ \bibinfo {pages}
  {173501} (\bibinfo {year} {2005})}\BibitemShut {NoStop}%
\end{thebibliography}%
\end{document}